\documentclass[a4paper,fleqn]{cas-dc}
\usepackage[applemac]{inputenc}
\usepackage{amsmath}
\usepackage{natbib}

\newcommand{\te}[1]{10^{#1}}
\newcommand{\ut}[1]{\hspace{1mm}\mathrm{#1}}
\newcommand{\vc}[1]{\boldsymbol{\mathrm #1}}
\newcommand{\vecu}{\vc{u}} 
\newcommand{\vecB}{\vc{B}}

\newcommand{\ddp}[2]{\displaystyle \dfrac{\displaystyle \partial #1}{\displaystyle \partial #2}}
\newcommand{\ddt}[2]{\displaystyle \dfrac{\displaystyle {\mathrm d} #1}{\displaystyle {\mathrm d} #2}}

\newcommand{\cit}[1]{\let\temp=\\ \centering #1\let\\=\temp}
\newcommand{\rev}[1]{{#1}}

\newcolumntype{L}[1]{>{\raggedright\let\newline\\\arraybackslash\hspace{0pt}}m{#1}}
\newcolumntype{C}[1]{>{\centering\let\newline\\\arraybackslash\hspace{0pt}}m{#1}}
\newcolumntype{R}[1]{>{\raggedleft\let\newline\\\arraybackslash\hspace{0pt}}m{#1}}

\def\tsc#1{\csdef{#1}{\textsc{\lowercase{#1}}\xspace}}
\tsc{WGM}
\tsc{QE}

\begin{document}
\let\WriteBookmarks\relax
\def\floatpagepagefraction{1}
\def\textpagefraction{.001}

\shorttitle{Kinematic control of geomagnetic reversals}
\shortauthors{Aubert et al.}

\title[mode = title]{Core-surface kinematic control of polarity reversals in advanced geodynamo simulations}

\author[1]{Julien Aubert}
\cormark[1]
\ead{aubert@ipgp.fr}
\author[1]{Maylis Landeau}
\author[1]{Alexandre Fournier}
\author[1]{Thomas Gastine}

\affiliation[1]{organization={Université Paris Cité, Institut de physique du globe de Paris, CNRS}, postcode={F-75005},city={Paris}, country={France}}


\maketitle

\begin{abstract}
The geomagnetic field has undergone hundreds of polarity reversals over Earth's history, at a variable pace. In numerical models of Earth's core dynamics, reversals occur with increasing frequency when the convective forcing is increased past a critical level. This transition has previously been related to the influence of inertia in the force balance. Because this force is subdominant in Earth's core, concerns have been raised regarding the geophysical applicability of this paradigm. Reproducing the reversal rate of the past million years also requires forcing conditions that do not guarantee that the rest of the geomagnetic variation spectrum is reproduced. These issues motivate the search for alternative reversal mechanisms. Using a suite of numerical models where buoyancy is provided at the bottom of the core by inner-core freezing, we show that the magnetic dipole amplitude is controlled by the relative strength of subsurface upwellings and horizontal circulation at the core surface. A relative weakening of upwellings brings the system from a stable to a reversing dipole state. This mechanism is purely kinematic because it operates irrespectively of the interior force balance. It is therefore expected to apply at the physical conditions of Earth's core. Subsurface upwellings may be impeded by stable stratification in the outermost core. We show that with weak stratification levels corresponding to a nearly adiabatic core surface heat flow, a single model reproduces the observed geomagnetic variations ranging from decades to millions of years. \rev{In contrast with} the existing paradigm, reversals caused by this stable top core mechanism become more frequent when the level of stratification increases i.e. when the core heat flow decreases. This suggests that the link between mantle dynamics and magnetic reversal frequency needs to be reexamined.
\end{abstract}

\section{\label{intro}Introduction}
The temporal variations of Earth's internal magnetic field span a wide range of time scales, from geomagnetic jerks occurring over years and less, to variations of the dipole reversal frequency taking place over hundred million years and longer \citep[see e.g.][for recent reviews]{Finlay2023,Domeier2023}. One of the challenges of geodynamo studies is to provide a holistic description of this signal across time scales. One goal in particular is to relate polarity reversals and their variability to the much shorter secular convective overturn, which is the fundamental time scale for Earth's core dynamics. Robust observational constraints are available within this range. Steadily improving observational reconstructions of geo- and paleomagnetic variations span centuries to millions of years \citep{Ziegler2011,Panovska2018,Nilsson2022,Constable2023}, and the reversal rate of the past 10 million years is also well documented \citep{Ogg2020}, with 5 major events over the past 2 Ma and 2-4 times as many excursions \citep{Laj2015Treatise} or incomplete reversals. This provides a rich quantitative basis against which models of Earth's core dynamics can be compared \citep[e.g.][]{Sprain2019}. 

The successful simulation of geomagnetic polarity reversals \citep{Glatzmaier1995} kickstarted the exploration of the physical parameter space for numerical models of the geodynamo. Systematic surveys \citep{Kutzner2002,ChristensenAubert2006,Driscoll2009b,Olson2014} have shown that when the convective forcing is increased, the system transitions from a state with a dominant axial dipole to one where this component vanishes and the magnetic field becomes dominated by multipoles. In the vicinity of this transition, it is possible to obtain a non-vanishing axial dipole presenting occasional sign changes \citep[see e.g.][]{Wicht2010SSR,Christensen2011}, providing a reasonable first-order description of geomagnetic reversals. More detailed recent comparisons of simulated and natural reversals \citep{Sprain2019,Meduri2021} confirm this success but also show that this description does not explain all aspects of paleomagnetic variations. This discrepancy likely stems from the difficulty to produce both a strong dipole and strong fluctuations in a single model. 

A second problem is that the frequency of reversals increases steeply with convective forcing in simulations \citep{Driscoll2009b,Olson2014}. The range of admissible forcing for the reproduction of a given reversal rate is very narrow, and therefore not necessarily compatible with the reproduction of the rest of the geomagnetic spectrum, nor with constraints related to Earth's heat budget. To circumvent this, composite spectra can be built from several distinct models in order to cover a broadband geomagnetic spectrum \citep{Olson2012}. Though this is attractive if one wishes to overcome the numerical challenge of simulating short and long timescales together, fully understanding the connection between core convection and geomagnetic reversals still requires a single model operating in realistic forcing conditions and accounting for the whole spectrum. 

A third problem is that in most geodynamo models, the dipole-multipole transition is thought to link with a change of force balance inside the core, where the increase of convective forcing brings the inertial force on par with the magnetic force \citep{ChristensenAubert2006,Tassin2021,Nakagawa2022}. The ratio between magnetic and inertial forces is well represented by the magnetic to kinetic energy ratio $E_\mathrm{mag}/E_\mathrm{kin}=B^2/\rho\mu U^{2}$ \citep{Schwaiger2019}, where $\rho=1.1\times\te{4}\ut{kg.m^{-3}}$ is the core mean density, $\mu=4\pi\te{-7}\ut{H.m^{-1}}$ is the magnetic permeability, $U$ and $B$ are typical velocity and magnetic field amplitudes. Taking $U\approx 5\times\te{-4}\ut{m.s^{-1}}$ and $B\approx4\ut{mT}$ for Earth's core  \citep[e.g.][]{Finlay2023} leads to $E_\mathrm{mag}/E_\mathrm{kin}$ values of a few thousands. This shows that at the system scale, inertial forces are unlikely to approach the magnetic force for any realistic level of forcing. This raises strong concerns on the applicability of numerical model results to Earth's core conditions \citep{Tassin2021}. Although inertial forces can conceivably become important at much smaller length scales of about 100 m, it is implausible that they may have a macroscopic influence on the system \citep{Christensen2011}, because magnetic diffusion should homogenize the field at these scales. An alternative proposal is that the transition is misinterpreted and that it is the buoyancy, rather than inertial forces, that causes the shift from stable to reversing dynamos \citep{Sreenivasan2014}\rev{, or buoyancy-driven changes in the properties of magnetohydrodynamic waves not involving inertia \citep{Majumder2024}.} This is an attractive suggestion because unlike inertia, buoyancy enters the main force balance thought to hold in Earth's core \citep{Yadav2016PNAS,Schwaiger2019}. However, with most of the existing reversing numerical dynamos operating near $E_\mathrm{mag}/E_\mathrm{kin}\approx 1$ \citep{Tassin2021} or reaching this value during reversal events \citep{Nakagawa2022,TerraNova2024}, it has remained difficult to distinguish the role of buoyancy from that of inertia in models. Recently,  \cite{Jones2025} have shown that increasing the magnitude of buoyancy fluctuations enables reversing dynamos with $E_\mathrm{mag}/E_\mathrm{kin}$ significantly departing above unity, giving support to the proposal of \cite{Sreenivasan2014}. A detailed force balance analysis of these models remains to be performed in order to ascertain this conclusion. \rev{Along similar lines, \cite{Frasson2025} also obtain reversing dynamos with $E_\mathrm{mag}/E_\mathrm{kin}$ up to 10 (though this also falls back to about 2 during reversals) by enforcing excess equatorial heat flow at the top of an otherwise adiabatically cooling core.}

Another possibility, and a potential solution to the above problems, is that the origin of geomagnetic reversals is of kinematic rather than dynamic nature, i.e. unrelated to the interior force balance. The magnetic induction equation alone indeed contains enough complexity to produce polarity reversals even with an imposed flow structure, as can be done with kinematic dynamos \citep{Gubbins2008kin}. Low-dimensional dynamical systems mimicking the behaviour of the induction equation have also been successful at explaining some of the properties of geomagnetic reversals \citep{Petrelis2009}. From a modelling standpoint, this hypothesis is the most attractive because it disconnects the occurrence of reversals from the need to increase forcing, and this forcing can then be adjusted to match the constraints associated with the rest of the geomagnetic spectrum. But this requires to gain a better understanding of the conditions under which geomagnetic reversals can be kinematically produced. In this study, we show that convective upwellings beneath the core surface exert a kinematic control on the dipole strength, and that reversals can be obtained if the top of the core is stabilised against convection.

Exploring the observable consequences of a stable top core is also interesting from a geophysical standpoint because of its connections with Earth's heat budget. Estimates based on ab-initio computations favour high values $k\approx100\ut{W.m^{-1}.K^{-1}}$ for the core thermal conductivity \citep[e.g][]{Davies2015}, implying that the adiabatic heat flow out of the core is about $Q_\mathrm{ad}\approx 15\ut{TW}$. Because this is similar to the mantle-side estimates for the total core heat flow $Q_\mathrm{CMB}$ \citep[e.g.][]{Frost2022}, a subadiabatic situation with $Q_\mathrm{CMB}<Q_\mathrm{ad}$ is likely, in which case a convectively stable region exists at the top of the core. Previously, \cite{Gastine2020} have shown that the presence of a stable layer can smooth the surface magnetic field and increases its stability via an electromagnetic skin effect. Here, we show that this in fact crucially depends on how the buoyancy is distributed in the core.  When convection is driven from the bottom, as is the case while the inner core is freezing, a stable layer can destabilise the dipole instead. 

This manuscript is organised as follows: section \ref{model} introduces the numerical model. The results are presented in section \ref{results}, and discussed in the light of available data and core evolution scenarios in section \ref{discu}.

\section{\label{model}Models and methods}
\subsection{Numerical geodynamo model description and inputs}
The numerical model solves the Navier-Stokes, thermo-chemical convective density anomaly transport and magnetic induction equations within a spherical shell of thickness $D=r_{o}-r_{i}$ (\rev{with} $r_{i}/r_{o}=0.35$ as in Earth's core at present) rotating at an angular velocity $\Omega$ around an axis $\vc{e}_{z}$, and filled with an incompressible fluid of density $\rho$, electrical conductivity $\sigma$ and magnetic permeability $\mu$. The model equations may be found in \cite{Aubert2017}. A spherical coordinate frame $(r,\theta,\varphi)$ is defined with unit vectors $\vc{e}_{r}, \vc{e}_{\theta},\vc{e}_{\varphi}$. The Boussinesq and magnetohydrodynamic approximations are used. The unknowns are the velocity field $\vecu$, magnetic field $\vecB$ and density anomaly field $C$. Introducing the magnetic diffusivity $\eta=1/\mu\sigma$, viscous and thermochemical diffusivities $\nu$ and $\kappa$, the magnetic Ekman, magnetic and hydrodynamic Prandtl numbers are defined as
\begin{eqnarray}
E_\eta&=&\dfrac{\eta}{\Omega D^{2}},\\
Pm&=&\dfrac{\nu}{\eta},\\
Pr&=&\dfrac{\nu}{\kappa}.
\end{eqnarray}
Values of $E_{\eta}$ and $Pm$ are reported in supplementary Table 1, together will all dimensionless inputs, and all models use $Pr=1$. \rev{A subset of models is also presented in Table \ref{somemodels}.}

Several different buoyancy profiles are analysed in this study. The base configuration is set up as in \cite{Aubert2023} and driven by a homogeneous mass anomaly flux $F$ at the inner boundary, while this flux vanishes at the outer boundary. This represents a reference situation with light elements release from inner core freezing and adiabatic heat flow at the top of the core. The flux-based Rayleigh number is defined as
\begin{equation}
Ra_{F}=\dfrac{g_{o}F}{4\pi\rho\Omega^{3}D^{4}},\label{RaF}
\end{equation}
where $g_{0}$ is the gravity at the outer boundary. 

Models labelled with 'Bot' in supplementary Table 1 \rev{and Table \ref{somemodels}} start from this base situation and further control the convective stability near the outer surface by adding an adverse radial background density anomaly profile in a top layer of thickness $H=10-290\ut{km}$:
\begin{equation}
\ddp{C_{0}}{r} (r\ge r_{o}-H) = -\dfrac{N^2\rho}{g_{o}} \dfrac{(r-r_o+H)}{H},\label{Cgradient}
\end{equation}
where $N$ is the Brunt-Väisälä frequency at the external boundary. \rev{The average stabilising density anomaly in the layer is $\Delta\rho_\mathrm{st}=N^{2}\rho H / 2 g_{o}$.}

Models with label 'Het' do not involve an adverse density gradient but instead impose a laterally heterogeneous mass anomaly flux at the outer boundary of peak-to-peak local amplitude $\Delta f$, with pattern as in \cite{Aubert2008_2} i.e. derived from deep mantle seismic tomography \citep{Masters2000}, representing the thermal control from an heterogeneous mantle. The top of the shell is buoyantly neutral on spherical average ($N=0$) but regionally stabilised by the imposed lateral mass anomaly flux heterogeneities. The fluid is most stable beneath the large low shear velocity provinces of the lower mantle situated in the equatorial African and Pacific regions \citep[similarly to][]{Mound2019,Mound2023}. 

Models with label 'Sup' explore a situation where heat flow out of the core is superadiabatic and the top of the core is convectively unstable. These are driven by homogeneous mass anomaly fluxes $F$ and $F_{o}$ at the inner and outer boundary, respectively, following the implementation of \cite{Aubert2009}, with $F/(F+F_{o})=0.8$ i.e. with 20\% top-driven and 80\% bottom-driven forcing.

Finally, for comparison with \cite{Gastine2020}, models with label 'Vol' reproduce the buoyancy distribution of their study. A constant, destabilising (positive), spherically symmetric background density anomaly gradient $F/4\pi D^{2} \kappa$ is maintained throughout the shell. Here buoyancy is evenly distributed in the volume, in contrast with the other set-ups where it is mainly located at the bottom of the shell. 

The fluid shell representing the outer core is electromagnetically coupled at both boundaries with an axially rotating inner core of conductivity $\sigma$, and an axially rotating outer spherical shell of basal conductance $\te{-4}\sigma D$ representing the mantle. To reproduce the geomagnetic westward drift, the inner core and mantle are furthermore gravitationally coupled together using the formulation detailed in \cite{Aubert2013b,Pichon2016}, which involves a coupling constant $\Gamma\tau$. Stress-free and non-penetrating mechanical conditions are employed at both boundaries. By giving direct access to the free stream responsible for geomagnetic variations, this choice facilitates the diagnostic of their origin (see equation \ref{budget} below). We have checked that the surface flow and magnetic diffusion exposed by a stress-free condition indeed correspond to those obtained below the viscous boundary layer in corresponding cases with a no-slip outer boundary, and that the reversal mechanism is therefore not dependent on the choice of outer boundary condition. The relevance of stress-free boundaries for modelling the geodynamo in conditions relevant to Earth's core has also been documented in \cite{Aubert2017}.

\subsection{Outputs}
The outputs of 41 model cases are listed in supplementary Tables 1,2 in dimensionless and dimensional forms, respectively. \rev{Table \ref{somemodels} also presents a subset of outputs for selected models}.

\begin{table*}
\rev{\begin{tabular}{lrrrrrrrrrrrrr}
Label  & Reversals? & Buoyancy & $E_{\eta}$ & $Ra_{F}$ & $Pm$ & $\dfrac{N}{N_{0}}$ & $\dfrac{\Delta f}{f_{0}}$ & $Rm$ & $\dfrac{E_\mathrm{mag}}{E_\mathrm{kin}}$ & $D_{12}$ & $\chi^{2}$ & $Q_\mathrm{PM}$  \\[0.3cm]
\hline\\[-0.3cm]
Neutral top & no & Bot & $6\times\te{-6}$ & $2.7\times\te{-5}$ & 5 & 0 & 0 & 1762 & 2.50 & 0.60 & 1.0 & 5.6 \\
Stable top & yes & Bot & $6\times\te{-6}$ & $2.7\times\te{-5}$ & 5 & 282.1 & 0 & 1816 &  1.89 & 0.47 & 2.2 & 3.9  \\
Stable top 0\% & yes & Bot & $7.5\times\te{-6}$ & $2.7\times\te{-5}$ & 4 & 325.7 & 0 & 1490 & 1.53 &  0.50 & 1.7 & 4.1\\
Stable top 29\% & yes & Bot & $7.5\times\te{-7}$ & $2.7\times\te{-7}$ & 0.4 & 325.7 & 0 & 1745 & 12.30 & 0.49 & 3.0 & 5.2 \\
Tomographic & yes & Tom & $6\times\te{-6}$ & $2.7\times\te{-5}$ & 5 &0 & 1.42 &  1831 &  1.67 & 0.45 & 3.5 & 5.8 \\
Unstable top & no & Sup & $6\times\te{-6}$ & $5.4\times\te{-5}$ & 5 & 0 & 0 & 2308 & 1.88 & 0.60 & 0.6 & 6.5 \\
\hline
\end{tabular}}
\caption{\label{somemodels}\rev{Subset of dimensionless input and output parameters for selected numerical models. All models have $Pr=1$. See text for definitions and Supplementary Tables 1,2 for the complete dataset.}}
\end{table*}

We monitor the root-mean squared and time averaged velocity $U$, magnetic field amplitude $B$, and convective power $p$ per unit volume \citep[as defined in][]{Aubert2017} in the shell. \rev{A typical convective density anomaly can then be obtained as $\Delta\rho_\mathrm{conv}=p/g_{o}U$.} The magnetic Reynolds number measuring the ratio of magnetic induction and diffusion is $Rm=\mu U D \sigma$. Because we explore a hypothesis of high thermal conductivity, this also implies a high electrical conductivity $\sigma\approx\te{6}\ut{S.m^{-1}}$ \citep{Davies2015}. Our models therefore explore values $Rm=950-2100$ that are typically higher than in previous studies \citep{Olson2017,Christensen2018,Nakagawa2022,Buffett2023,TerraNova2024}.

The kinetic and magnetic energy densities are $E_\mathrm{kin}=\rho U^{2}/2$ and $E_\mathrm{mag}=B^{2}/2\mu$, respectively. The surface flow $\vecu({r_o})$ is expressed using the following decomposition:
\begin{equation}
\vc{u}(r_{o})=\left(\dfrac{1}{\sin \theta} \ddp{T}{\varphi}+\ddp{S}{\theta}\right)\vc{e}_{\theta} +  \left(-\ddp{T}{\theta}+\dfrac{1}{\sin \theta} \ddp{S}{\varphi}\right)\vc{e_\varphi}.\label{torspher}
\end{equation}
The toroidal and spheroidal scalars $T$ and $S$ respectively representing the non-divergent and divergent (upwelling) parts of the surface flow are constantly recorded up to spherical harmonic degree 30. Using the continuity condition of the incompressible fluid, the surface upwelling relates to the spheroidal scalar $S$ through
\begin{equation}
\ddp{u_{r}}{r}(r_{o})= -r_{o} \nabla^2_{H} S.
\end{equation}
where $u_{r}=\vecu\cdot\vc{e}_{r}$ and $\nabla^2_{H} $ is the horizontal part of the Laplace operator. From this recording we respectively define the surface circulation $U_\mathrm{surf}$ and upwelling strength $W$ as the root-mean-squared and time-averaged values of $\vc{u}$ and $\partial u_{r}/\partial r$ over the outer surface of the model. 

Our analysis compares the time scale for surface circulation defined as $\tau_\mathrm{surf}=D/U_\mathrm{surf}$ with a time scale typical of magnetic flux expulsion by subsurface upwellings. \rev{To evaluate flux expulsion, we consider the magnetic boundary layer present beneath the outer surface of the fluid shell, where the toroidal magnetic field induced in the interior needs to accomodate a near-vanishing value at $r=r_{o}$. The thickness $\delta$ of this layer can be obtained by equating the induction and diffusion terms in the magnetic induction equation, leading to a  typical scaling $\delta\sim Rm^{-1/2}D$ \citep[e.g.][]{TerraNova2020}. A precise determination can be obtained by directly computing the integral length scale present in the magnetic diffusion term i.e. by using the magnetic dissipation length scale as defined in \cite{Aubert2017}. In the following we take this second route.} Consistently with the time needed to reach steady state in kinematic models of flux expulsion \citep{Troyano2020}, we \rev{then write the  time scale for magnetic flux expulsion} as $\tau_\mathrm{exp}=D/W\delta$. 

The axial dipole amplitude $g_{1}^{0}$ is defined using the classical Gauss coefficient formulation \citep{Alken2021}, with reference at the surface radius $r_\mathrm{E}=6371.2\ut{km}$ of the Earth. The corresponding axial dipole moment is $4\pi r_\mathrm{E}^{3}~|g_{1}^{0}|/\mu$. The dipole colatitude $\theta_{d}=\cos^{-1} \left(g_{1}^{0}/\sqrt{(g_{1}^{0})^2+(g_{1}^{1})^2+(h_{1}^{1})^2}\right)$ and latitude $\lambda_{d}=\pi/2-\theta_{d}$ additionally involve the equatorial dipole Gauss coefficients $g_{1}^{1}$ and $h_{1}^{1}$.

The morphological similarity of the model output with the geomagnetic field of the past centuries is diagnosed through a set of compliance criteria defined in \cite{Christensen2010}. In Supplementary table 1 \rev{and Table \ref{somemodels}} we report on the final rating $\chi^{2}$ that summarises the departure of these criteria from target Earth values, with excellent compliance being characterised by $\chi^{2}\le 2$ and good compliance by $\chi^{2}\le 4$. 

The temporal variability of the output is characterised using the standard master secular variation time scale $\tau_\mathrm{SV}^{1}$ \citep{Holme2011,Lhuillier2011b}. This is obtained by computing, at any altitude above the model surface, the time-averaged spectra $\left<B^2(\ell)\right>$ and $\left<\dot{B}^2(\ell)\right>$ of the magnetic field and its temporal variations as functions of the spherical harmonic degree $\ell$, and then adjusting by least-squares fitting a functional form $\tau_\mathrm{SV}^{1}/\ell$ to $\tau_\mathrm{SV}(\ell)=\displaystyle\sqrt{\left<B^2(\ell)\right>/\left<\dot{B}^2(\ell)\right>}$ between degrees $\ell=2$ and $\ell=13$. The range of compliance with recent geomagnetic variations is $\tau_\mathrm{SV}^{1}=370-470\ut{yr}$ \citep{Lhuillier2011b}.

The variability of the output magnetic field over time scales of millions of years is first diagnosed with the number of reversals $\mathrm{\# R}$ observed in each model sequence. The label 'N/A' in supplementary Table 1 refers to a situation where the axial dipole is too weak to characterise polarity reversals. The quantity $\tau_\mathrm{trans}$ is the ratio between the time spent by the axial dipole in a transitional state and the entire duration of the model sequence, a transitional state being defined by $|\lambda_{d}|<45^{o}$. 

Also computed are a set of paleomagnetic quantities and compliance diagnostics defined in \cite{Sprain2019}. The two 'Model G' coefficients $a$ and $b$ characterize the dependence of paleosecular variation with site latitude \citep{McFadden1988}. The diagnostic $I$ is the inclination anomaly of the time averaged output field respectively to a geocentric axial dipole. The diagnostic $V\%$ is the ratio between the interquartile range and median of the virtual dipole moment distribution. Associated to these diagnostics are four measures $Qa,Qb,QI,QV\%$ representing the deviation of diagnostics from target Earth values for the past 10 Ma \citep{Sprain2019}. A fifth measure characterises the position of $\tau_\mathrm{trans}$ with respect to the interval $0.0375-0.15$ deemed Earth-like in \cite{Sprain2019}. The quantity $\Delta Q_\mathrm{PM}$ is the sum of these five measures, and an integer score $Q_\mathrm{PM}$ provides the number of passed criteria. The range of compliance with paleomagnetic variations is $\Delta Q_\mathrm{PM}\le 5$. 

At the core surface, we also report on the dipolarity levels $D_{4}$ and $D_{12}$ \citep[with this latter quantity also termed $f_\mathrm{dip}$ in][]{ChristensenAubert2006,Nakagawa2022,TerraNova2024} representing the mean ratio between the root-mean-squared dipole (including axial and equatorial parts) and total magnetic field amplitudes up to spherical harmonic degree 4 and 12, respectively. The estimated ranges of compliance for the past 10 Ma are $D_{4}=0.58-0.94$ and $D_{12}=0.36-0.72$ \citep{Meduri2021}. At the Earth surface, we report on the median value $(AD/NAD)_\mathrm{surf}$ of the ratio between axial dipole and components other than the axial dipole up to degree 10. The estimated range of compliance is $(AD/NAD)_\mathrm{surf}=4.8-26.4$ \citep{Biggin2020}. 

The analysis is complemented with the paleosecular variation index of \cite{Panovska2017} characterising the combined deviation of virtual geomagnetic poles away from the rotation axis, and of the virtual dipole moment away from the reference present value. The asymmetry in growth and decay of the dipole moment is diagnosed using the definitions and procedure of \cite{Buffett2023}. Dipole moment time series are binned into non-overlapping temporal windows of given duration (ranging from 1 to 50 kyr) and the trend within each window is evaluated by least-squares fitting. The skewness of the resulting distribution of trends is then computed together with its standard error range as a function of window duration.  

\subsection{Dimensioning}  
Dimensionless quantities in Supplementary table 1 \rev{and Table \ref{somemodels}} are presented relative to the shell thickness $D$, the magnetic diffusion time $\tau_{\eta}=D^{2}/\eta$, the Elsasser magnetic field unit $B_{0}=\sqrt{\rho\mu\eta\Omega}$, the volumetric power unit $p_{0}=\rho\eta^{2}\Omega/D^{2}$, the buoyancy frequency unit $N_{0}=\sqrt{\Omega\eta}/D$ and the unit $G_{0}=\rho D^{3}\eta$ for the gravitational coupling constant. The choice of these units is underlain by the path theory \citep{Aubert2017,Aubert2023}, through which dimensional values relevant to Earth's core can be obtained. Along suitably chosen paths in model parameter space leading to the physical conditions of Earth's core, the procedure fundamentally rests on the invariance of the leading-order quasi-geostrophic, magneto-Archimedes-Coriolis force balance that governs the dynamics on time scales comparable to or longer than the secular core overturn time $D/U$. Dimensionless quantities are therefore dimensioned for the conditions of Earth's core according to the preservation of this balance. For each model case, this provides a well-defined dimensional physical equivalent from the standpoint of long-term core dynamics, that can be accurately compared to observations \citep[with e.g. a typical 10\% uncertainty for the magnetic field amplitude,][]{Aubert2017}. \rev{The energetics of the resulting dimensional system are also directly comparable to the power budget of Earth's core.}

To describe the Earth's core conditions, we use the constants $D=2260\ut{km}$, $\Omega=2\pi/(1 \ut{day})$, $\rho=11000\ut{kg/m^{3}}$, $g_{o}=10\ut{m.s^{-2}}$ and $\mu=4\pi\te{-7}\ut{H.m^{-1}}$. Each model is characterised by an integer position $\beta$ on a logarithmic scale along a parameter space path of 7 decades connecting this model to the conditions of the core. At the end of this path (EOP), the magnetic Ekman number relevant to Earth is \citep{Aubert2017}:
\begin{equation}
E_{\eta} \mathrm{(EOP)}=\sqrt{\te{(\beta-7)}} E_{\eta} \mathrm{(model)}
\end{equation}
The dimensional magnetic diffusivity $\eta$ and core conductivity $\sigma$ are then obtained through $\eta=1/\mu\sigma=\Omega D^2  E_{\eta} \mathrm{(EOP)}$. Using this and the above supplied constants, we obtain the dimensioning units $D$, $\tau_{\eta}$, $B_{0}$, $p_{0}$, $N_{0}$ and $G_{0}$ needed to cast the dimensionless values of supplementary Table 1 into their dimensional equivalents at Earth's core conditions reported in supplementary Table 2. This table in particular reports the total power $P$ obtained by multiplying the dimensional volumetric power $p$ with the outer core volume. The number of reversals $\mathrm{\# R}$ obtained in each model sequence is also converted into a dimensional reversal rate $R=\mathrm{\# R}/\tau_\mathrm{sim}$, where $\tau_\mathrm{sim}$ is the dimensional duration of the sequence. Within our dataset, there exists a good correlation between $R$ and the relative time $\tau_\mathrm{trans}$ spent in a transitional state, with a reversal being obtained on average after $10^{4}$ years of transitional time. Because transitional states are intrinsically more frequent than reversals, $\tau_\mathrm{trans}$ provides a more reliable estimate of the true reversal rate. The rates presented in the following are therefore those estimated from $\tau_\mathrm{trans}$. 

For cases in the Het set-up, the outer boundary mass anomaly flux heterogeneity $\Delta f$ is expressed in supplementary Table 1 \rev{and Table \ref{somemodels}} relatively to $f_{0}=F/4\pi r_{o}^{2}$. In supplementary Table 2, this can be converted into a dimensional heat flow heterogeneity $\Delta q=q_{0} \gamma \Delta f/f_{0}$, where the constant $\gamma=1.6$ and the average heat flow per unit surface $q_{0}=Q_\mathrm{CMB}/4\pi r_{o}^{2}=105\ut{mW.m^{-2}}$ of these cases are determined in appendix \ref{thermomodel}. 

Model cases in Supplementary tables 1,2 are grouped into sets Bot[1-3], Het, Sup and Vol, with each set targeting a specific end-of-path value of the core conductivity $\sigma$ and distribution of convective buoyancy sources. The set Bot3 contains two equivalent models computed at 0\% and 29\% of the same parameter space path, targeting the same end-of-path convective power, stable layer configuration and strength. Models within each set otherwise belong to different parallel paths characterised by different start-of-path conditions, in order to explore a range of possible final Earth values for the above parameters.

\subsection{Numerical implementation.} 
The numerical implementation involves a decomposition of the fields in spherical harmonics up to degree and order 133, and a discretisation in the radial direction on a second-order finite-differencing scheme over $NR$ grid points. We use the spherical harmonics transform library SHTns \citep{Schaeffer2013} available at {\tt https://bitbucket.org/nschaeff/shtns}. Time stepping is semi-implicit with second-order accuracy. To ensure the correctness and reproducibility of results, numerical solutions were also benchmarked against those  independently obtained from the MagIC simulation code \citep{Wicht2002} available at {\tt https://magic-sph.github.io}. Solutions at advanced positions $\beta=1,2$ along parameter space paths (14 and 29\%) are approximated with hyperdiffusion applied to the velocity and density anomaly fields, but not to the magnetic field which remains natively resolved. This approach is physically justified in \cite{Aubert2017} and validated against fully resolved simulations in \cite{Aubert2019b}. At spherical harmonic degrees $\ell$ larger than a cut-off $\ell_{h}$, the native diffusivities $\nu,\kappa$ are replaced by effective diffusivities $(\nu_\mathrm{eff},\kappa_\mathrm{eff})=(\nu,\kappa)q_{h}^{\ell-\ell_{h}}$, with the values of $q_{h}$ and $\ell_{h}$ listed in supplementary Table 1. The cut-off value $\ell_{h}=30$ is well separated from the typical length-scale $\ell\approx 10$ of convection in our models \citep{Schwaiger2021}. Supplementary Table 1 also lists the values of $NR$ and durations $\tau_\mathrm{sim}$ of model sequences, which range from 1.1 to 12 magnetic diffusion times $\tau_\eta$ depending on the observed stability of the axial dipole.

\subsection{Sources of axial dipole temporal variations}
Our analysis involves a budget of the contributions to the axial dipole temporal variations, which can be written as latitudinal integrals at the outer boundary ($r=r_{o}$) of the model \citep{Olson2006,Finlay2016}: 
\begin{equation}
\begin{split}
\left<\ddt{g_1^0}{t} \right >= 0 = \dfrac{1}{\pi}\int_{0}^{\pi}-\dfrac{3\pi r_{o}^{2}}{4 r_\mathrm{E}^{3}} \left<\overline{u_{\theta} B_r}\right> \sin^{2}\theta\,d\theta \\
+ \dfrac{1}{\pi}\int_{0}^{\pi} \eta\dfrac{3\pi r_{o}}{4 r_\mathrm{E}^{3}} \left<\ddp{\overline{B_{r}}}{\theta}-\ddp{r\overline{B_{\theta}}}{r}\right>\sin^{2}\theta\,d\theta.
\end{split}
\label{budget}
\end{equation}
Here $u_{\theta}=\vecu\cdot \vc{e}_\theta$, $B_{r}=\vecB\cdot \vc{e}_r$, $B_{\theta}=\vecB\cdot \vc{e}_\theta$, the angle brackets denote the time average as above, and the overbar denotes the azimuthal average. The first integral in the right-hand side is the contribution to dipole generation from induction by meridional surface flows. The dipole is for instance reinforced through poleward migration of radial magnetic flux of normal polarity. The second integral is the contribution from magnetic diffusion. The inductive contributions from upwelling (divergent) and non-divergent surface flows can furthermore be separated by using equation (\ref{torspher}) and substituting $u_\theta= \partial S/\partial \theta$ or $u_\theta= \partial T/(\sin \theta\, \partial \varphi)$ into equation (\ref{budget}), respectively. 

\section{\label{results}Results}
\subsection{Reversals in a stable top core}
\begin{figure*}
\centerline{\includegraphics[width=17.5cm]{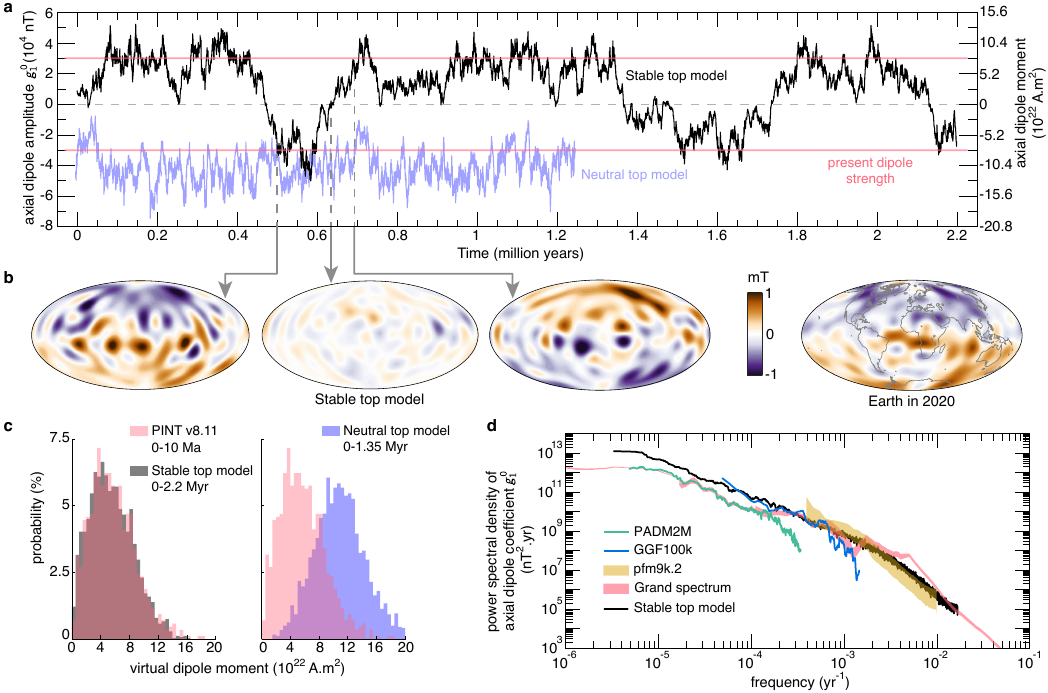}}
\caption{\label{gnmearth} a: Time series of the axial dipole amplitude $g_{1}^{0}$ obtained in the Stable top and Neutral top models. b: Hammer projections of the radial magnetic field at the core surface (filtered at spherical harmonic degree 13, orange is outwards) before, during and after a polarity reversal from the Stable top model, compared to the present-day structure from the International Reference Geomagnetic Field model \citep{Alken2021}. c: Histograms of the virtual dipole moment distributions from the Stable top and Neutral top models, downsampled following the procedure described in \cite{Sprain2019} to match a realistic distribution of paleomagnetic samples and locations, and compared to the distribution from the PINT paleointensity database \citep{Bono2022} filtered for the epoch range $0-10\ut{Ma}$ and determinations with quality index 3 or higher. d: Frequency-domain power spectral decomposition of the axial dipole time series from the Stable top model, compared to the paleomagnetic field models PADM2M \citep{Ziegler2011},  GGF100k \citep{Panovska2018}, the 95\% confidence range of model pfm9k.2 \citep{Nilsson2022} , and the $\pm 1$ standard deviation range of a grand spectrum obtained by compositing spectral estimates from data sets covering different frequency ranges \citep{Constable2023}.}
\end{figure*}

In the reference model, the Neutral top case (\rev{Table \ref{somemodels},} supplementary Tables 1,2), we explore a situation where the top of the core is neutrally buoyant, as is the case if $Q_\mathrm{CMB}$ exactly matches the adiabatic value $Q_\mathrm{ad}$. The convective power of this model is adjusted to $P\approx3\ut{TW}$, a plausible present value in the light of core evolution scenarios \citep{Labrosse2015,Nimmo2015evolution} that additionally enables an Earth like magnetic variation time scale $\tau_\mathrm{SV}^{1}=455\ut{yr}$. This produces a non-reversing magnetic dipole of strength exceeding that of the present geomagnetic field at most times (Fig. \ref{gnmearth}a). Attempting to obtain reversals by further increasing $Q_\mathrm{CMB}$ above $Q_\mathrm{ad}$ (superadiabatic cases 35 and Unstable top), as done classically, is not adequate here because $\tau_\mathrm{SV}^{1}$ becomes too short and gets outside the Earth-like range 370-470 yr before reversals can be obtained. This shows that at the parameters where our models operate (particularly the low magnetic Ekman number $E_\eta=6\times\te{-6}$), the convective power level needed to obtain reversals by forcing is already too high to quantitatively account for the observed secular geomagnetic variations. \rev{The capping of forcing set by the magnetic variation time scale $\tau_\mathrm{SV}^{1}$ should pertain to Earth's core conditions because this time scale is about invariant along parameter space paths of decreasing $E_\eta$ towards these conditions \citep{Aubert2018}.}

\begin{figure*}
\centerline{\includegraphics[width=12cm]{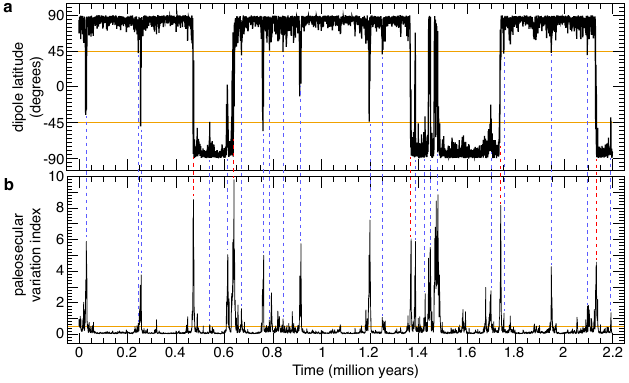}}
\caption{\label{PSVindex} Time series of (a) the dipole latitude $\lambda_{d}$, (b) the average paleosecular variation index \citep{Panovska2017} over sites evenly distributed at the Earth surface, in the temporal sequence from the Stable top model. Reversals (red dashed vertical lines) and excursions (blue dashed lines) are detected with  dipole latitudes less than 45 degrees away from the equator (orange lines in panel a) and spikes in the paleosecular variation index exceeding the value 0.5 (orange line in panel b).}
\end{figure*}

We next enforce a relatively weak stable region atop the core (Stable top model, Fig. \ref{gnmearth}a), as is the case if $Q_\mathrm{CMB}$ becomes slightly subadiabatic. Rather unexpectedly, the dipole then undergoes reversals sharing a high level of similarity with observations. Five complete reversal events are obtained over $2.2\ut{Myr}$, a rate similar to Earth in the past 2 Ma, as well as 21 excursions, consistent with the rate observed during the Brunhes \citep[][see Fig. \ref{PSVindex} for a precise identification of these events]{Ogg2020,Laj2015Treatise}. Fig. \ref{gnmearth}a also shows that during phases of stable polarity, the dipole amplitude matches the present-day observed value. The magnetic field at the core surface is also morphologically similar to Earth at present (Fig. \ref{gnmearth}b, and $\chi^{2}=2.2$), with westward-drifting equatorial flux patches of normal polarity as well as high-latitude lobes of concentrated flux, two features well identified in observations \citep{Finlay2023}. \rev{Models with uniform boundaries tend to favour a single north/south pair of lobes, somehow different from the dual pair seen for Earth at present. This single pair is anchored to a spontaneously arising eccentric interior flow gyre \citep[similar to][]{Schaeffer2017}, that is seen in Earth \citep{Finlay2023} at present, and periodically over the past millenia \citep{Suttie2025}.} The magnetic variation time scale $\tau_\mathrm{SV}^{1}=457\ut{yr}$ is unaffected by the presence of the stable top region and remains in line with present-day estimates. The morphological similarity with paleomagnetic variations from the past 10 Ma is attested by reproducing the observed distribution of virtual dipole moments (Fig. \ref{gnmearth}c) and by reasonable adherence to the paleomagnetic semblance criteria of \cite{Sprain2019} (supplementary Table 1). Here this model achieves $\Delta Q_\mathrm{PM}=3.9$, $Q_\mathrm{PM}=3$. \rev{Though the reversal rate of is similar to Earth in the past 2 Ma, the transitional time criterion is not matched because this criterion is calibrated with the higher rate of the past 10 Ma. The most significant discrepancy with the paleomagnetic field is therefore the coefficient $a=15^o$, reflecting a long-term variability of the field higher than the typically observed value $a\approx 10^{o}-13^{o}$  \citep{Sprain2019,Biggin2020}.} The frequency-domain distribution of axial dipole variations matches existing observational reconstructions from decades to tens of millenia (Fig. \ref{gnmearth}d). At longer time scales, the model overpredicts variations reconstructed from sedimentary records by a constant factor 3-4 in power (1.5 to 2 in variance). This difference is expected because this reconstruction also underrepresents the variance of absolute paleointensity determinations by a similar factor \citep{Ziegler2011}. 

\subsection{Top-down kinematic control of reversals}

\begin{figure}
\centerline{\includegraphics[width=8cm]{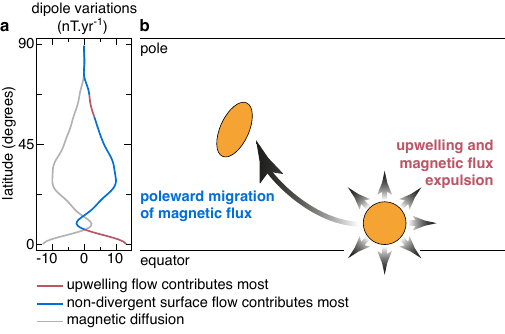}}
\caption{\label{schematics} a: Latitudinal profiles of the time-averaged contributions from core surface flow and magnetic diffusion to the variation $\mathrm{d}g_{1}^{0}/\mathrm{d}t$ of the axial dipole, in the Neutral top model. Presented are the integrands in equation (\ref{budget}), obtained from a 177 kyr-long subset during which the surface flow, the magnetic field and its gradients have been recorded at the native resolution of the model. Contributions constructive to dipole generation are represented positive by convention. The color-coding of the inductive profile separates regions of dominant contributions from upwelling (divergent) and non-divergent surface flows. b: Schematic description of the two corresponding kinematic processes acting on core surface magnetic flux patches over an hemisphere and sustaining the dipole against magnetic diffusion.}
\end{figure}

To understand the relation between convective stability and polarity reversals, we examine in Fig. \ref{schematics}a the sources of variations in the axial dipole amplitude at the surface of the core, as computed from equation (\ref{budget}). There, we find that magnetic diffusion acts against the generation of the dipole while, on time average, induction by core surface flow provides a constructive contribution that equilibrates the budget. The inductive part can furthermore be broken into contributions from a divergent (upwelling) and non-divergent (surface circulation) surface flows. Among these two contributions, only the divergent flow can create new magnetic energy at the core surface \citep{Huguet2018}. Near the equator (Fig. \ref{schematics}b), this manifests as newly created magnetic flux patches within upwellings, of normal polarity similar to the pre-existing dipole. The non-divergent part is the planetary-scale surface gyre flow \citep[see e.g.][]{Finlay2023} which subsequently transports these patches towards the poles. We anticipate that the system becomes prone to reversals when the magnetic field is advected away towards the pole before it has had sufficient time to build up near the equator. \rev{This suggests that the relative position of the time scales $\tau_\mathrm{exp}$ for subsurface magnetic flux expulsion and $\tau_\mathrm{surf}$ for surface circulation plays an important role in determining the level of the dipole and its reversal properties.}

\begin{figure}
\centerline{\includegraphics{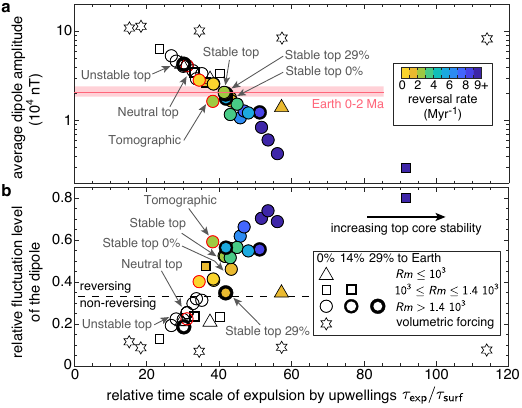}}
\caption{\label{revstats} a: Time-averaged absolute axial dipole coefficient $\left<|g_{1}^{0}|\right>$ as a function of the time scale ratio $\tau_\mathrm{exp}/\tau_\mathrm{surf}=U_\mathrm{surf}/W\delta$ between magnetic flux expulsion by subsurface upwellings and surface circulation. An estimate of $\left<|g_{1}^{0}|\right>$ for Earth in the past 2 Ma is reported \citep[pink line, from][]{Ziegler2011}, together with a range (shaded region) corresponding to averages computed over the Brunhes (0-780 ka) and earlier (780 ka - 2 Ma) epoch ranges.  {\bfseries b,} Standard deviation of $|g_{1}^{0}|$ normalised by its time-average value. The symbol color coding indicates an estimate of the reversal rate, with empty symbols denoting non-reversing cases. Symbol shapes indicate the ranges of magnetic Reynolds number $Rm$ for bottom-driven cases (labels Bot, Het and Sup in supplementary Tables 1,2), and stars denote cases with volumetric forcing (label Vol). Symbol rim thicknesses indicate the progression of models towards Earth's core conditions along paths in parameter space \citep{Aubert2017}. Red rims indicate cases with heterogeneous thermal control from the mantle.}
\end{figure}

This interpretation is supported by a systematic survey where the depth and strength of the stable top region is varied, \rev{in addition to the dynamo power, buoyancy distribution and electrical conductivity. The ratio $\tau_\mathrm{exp}/\tau_\mathrm{surf}=U_\mathrm{surf}/W\delta$ is more sensitive to the effect of increasing top core stability, which decreases the upwelling $W$ at the top of the core (Supplementary Fig. 1) while keeping the surface circulation $U_\mathrm{surf}$ largely unchanged, than to variations in dynamo power which affect $U_\mathrm{surf}$ and $W$ in a similar way. Fig. \ref{revstats}a shows that} the time-average amplitude of the axial dipole monotonically decreases \rev{as $\tau_\mathrm{exp}/\tau_\mathrm{surf}$ increases and subsurface upwellings are weakened by an increasing top core stability}. Dipole fluctuations are comparatively less affected, and their level increases relative to the average \rev{as $\tau_\mathrm{exp}/\tau_\mathrm{surf}$ increases} (Fig. \ref{revstats}b). Reversals are found to occur when fluctuations exceed a third of the average. Earth’s estimated average dipole strength over the past 2 Ma \citep {Ziegler2011} coincides with that obtained in our models at a realistic reversal rate of 2-3/Myr (Fig. \ref{revstats}a). The relative fluctuation level also decreases with a decreasing magnetic Reynolds number, such that reversals caused by a stable top core become significantly less frequent or even disappear in our cases with $Rm<1000$ (triangles in Fig. \ref{revstats}b).

In contrast with these results, similar stable top layers were previously found to damp magnetic fluctuations and to stabilise the dipole \citep{Christensen2018,Gastine2020}. Reproducing the results from \cite{Gastine2020}, we found \rev{(Supplementary Fig. 2)} that the shallow convective forcing imposed by an even distribution of buoyancy in the volume turns surface magnetic diffusion into a constructive net contributor to the generation of the dipole in equation (\ref{budget}). This suppresses the sensitivity of the dipole to the strength of upwelling (see \rev{pentagrams} in Fig. \ref{revstats}a,b). \cite{Christensen2018} used bottom-driven convection but magnetic Reynolds numbers well below 1000, which also suppresses reversals. Our results therefore stem from the combined use of bottom-driven convection and high magnetic Reynolds numbers, two factors pertaining to Earth's core under the hypothesis of high core thermal and electrical conductivities.

\begin{figure}
\centerline{\includegraphics[width=8.5cm]{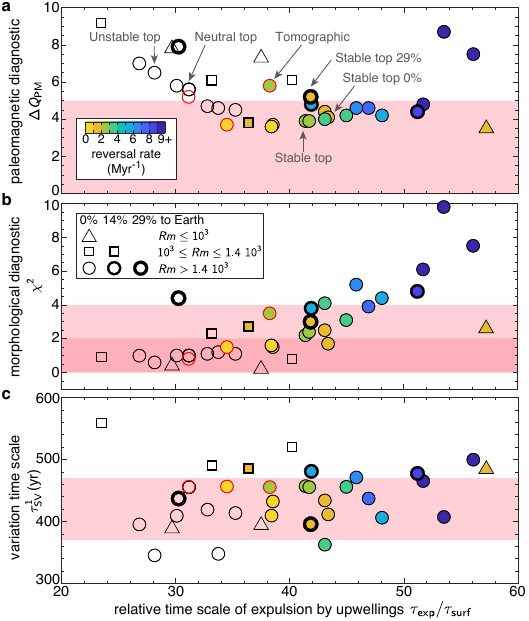}}
\caption{\label{compliance} Plotted against the time scale ratio $\tau_\mathrm{exp}/\tau_\mathrm{surf}$ between expulsion by subsurface upwelling and surface circulation are the model diagnostics a: $\Delta Q_\mathrm{PM}$ for paleomagnetic variations \citep{Sprain2019}, b: $\chi^{2}$ for the morphology of the core surface field \citep{Christensen2010}, and c: the master secular variation time scale \citep{Lhuillier2011b} $\tau_\mathrm{SV}^{1}$. Coding of symbols is the same as in Figure \ref{revstats}. Shaded regions in (a-c) delineate the ranges of compliance with the geo- and paleomagnetic field.} 
\end{figure}

\subsection{Compliance with geomagnetic observations}
The systematic survey also shows that similarly to the Stable top model presented in Figure \ref{gnmearth}, several cases with weakened upwellings reproduce the morphology and variations of the present and past geomagnetic field. \rev{For each of these cases, the dataset includes a corresponding non-reversing case with the same level of convective forcing but a neutral top, showing that it is indeed the inclusion of a stable top that is responsible for reversals, and not the classical forcing-driven paradigm.} The best compliance with paleomagnetic variations from the past 10 Myr is obtained for $\tau_\mathrm{exp}/\tau_\mathrm{surf}=35-45$ (diagnostic $\Delta Q_\mathrm{PM}$ in Fig. \ref{compliance}a), the range that also enables a correct reproduction of the average paleomagnetic dipole (Fig. \ref{gnmearth}a). \rev{An increase of $\tau_\mathrm{exp}/\tau_\mathrm{surf}$ simultaneously increases the paleomagnetic coefficient $a$ and the transitional time $\tau_\mathrm{trans}$, and the main residual source of discrepancy is the difficulty \citep[previously also noted by][]{Meduri2021} to achieve an Earth-like value for both in the same model (Supplementary Fig. 3). Examining the dominance of the dipole with respect to other magnetic field components, we nevertheless notice that the stable top core mechanism can produce reversals while maintaining dipolarity values close to (or even above) $D_{12}=0.5$, typically higher than those previously obtained by the forcing-driven paradigm \citep{Meduri2021}. We are incidentally also able to confirm the relationship found by these authors between $D_{12}$ and $\Delta Q_\mathrm{PM}$ (Supplementary Fig. 4), with the best paleomagnetic compliance (lowest $Q_\mathrm{PM}$) being obtained when $0.4\le D_{12} \le 0.6$. Fig. \ref{compliance}b,c furthermore shows that in the range $\tau_\mathrm{exp}/\tau_\mathrm{surf}=35-45$},  the morphology of the surface magnetic field (diagnostic $\chi^{2}$) as well as its variation time scale $\tau_\mathrm{SV}^{1}$ are both Earth-like. Controlling the strength of subsurface upwellings therefore enables a joint reproduction of the Earth's magnetic field from short (historical, secular) time scales to long, paleomagnetic time scales, as well as its in-between spectral content (Figure \ref{gnmearth}d). 

\begin{figure*}
\centerline{\includegraphics[width=17.5cm]{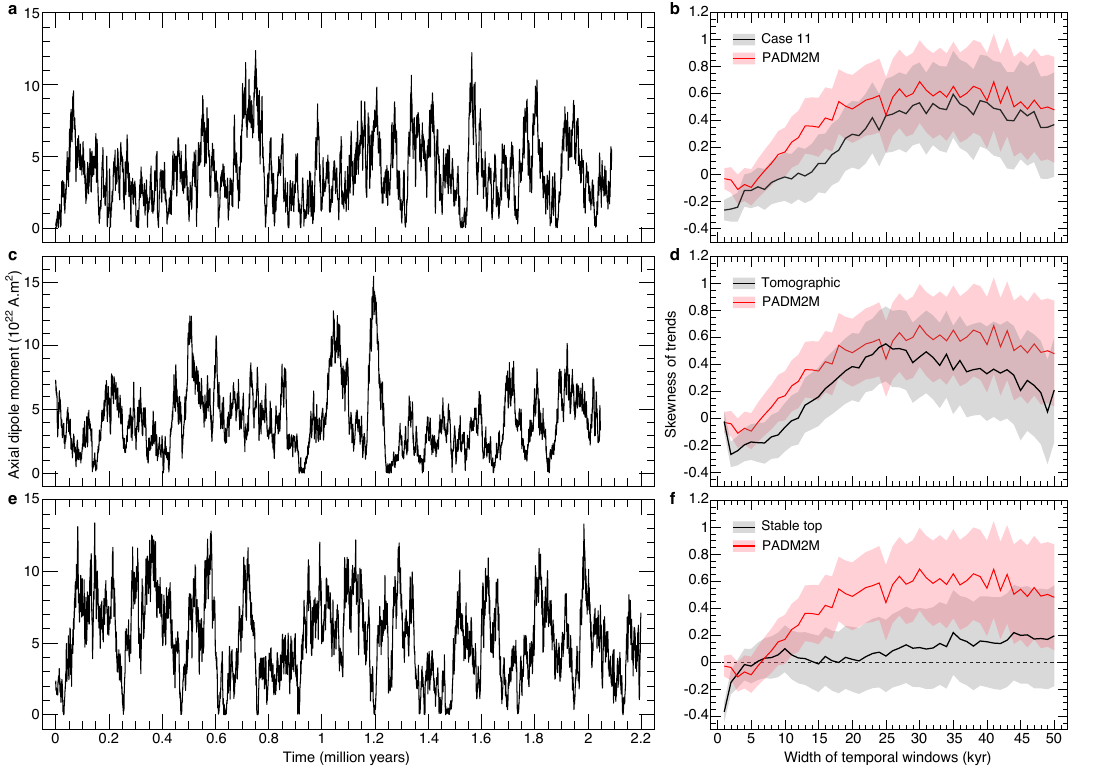}}
\caption{\label{skewness} a,c,e: Time series of axial dipole moments in model cases 11, Tomographic and Stable top (supplementary Tables 1,2). b,d,f: Skewness of trends in the dipole moment time series \citep[computed as in][]{Buffett2023}, plotted as a function of the width of the temporal window over which trends are evaluated, compared to the output from the PADM2M paleomagnetic field mode \citep{Ziegler2011}. Shaded areas represent the $\pm 1$ standard error range on the determination of skewness \citep[determined as in][]{Buffett2023} .}
\end{figure*}

The mechanism described in Fig. \ref{schematics} suggests different causes, and different time scales for the inductive creation of the dipole and its diffusive destruction. We therefore expect a general trend of slow decay and rapid recovery of the field, as has been suggested from the million-year dipole variation record \citep{Valet1993,Ziegler2011b,Buffett2023}. While this was not systematically observed in all the realisations of our numerical models, several of our cases indeed showed a positively skewed distribution of dipole variation trends (Fig. \ref{skewness}) matching the observed distribution.

\subsection{Relevance to Earth's core}
\begin{figure}
\centerline{\includegraphics[width=8.5cm]{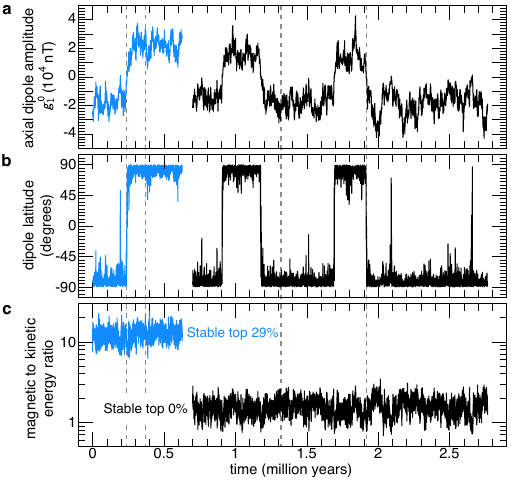}}
\caption{\label{pathrev} Time series of (a) the axial dipole coefficient $g_{1}^{0}$, (b) the dipole latitude $\lambda_{d}$, and (c) the interior magnetic to kinetic energy ratio $E_\mathrm{mag}/E_\mathrm{kin}$ obtained in the Stable top 0\% and 29\% models, which only differ by their distance to Earth's core conditions along the same parameter space path. Force balances in the interior of the core are examined in Fig. \ref{forcebal} at the times marked by vertical dashed lines.}
\end{figure}

\begin{figure*}
\centerline{\includegraphics[width=17.5cm]{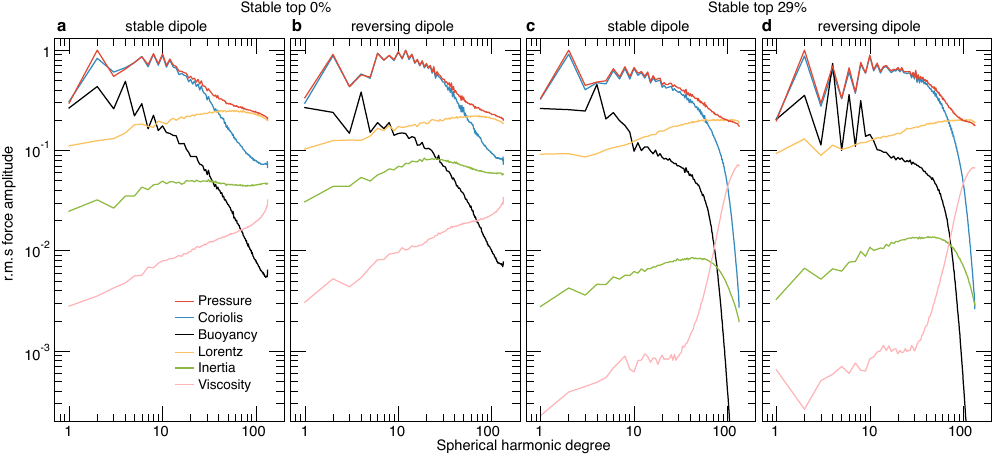}}
\caption{\label{forcebal}Root-mean-squared amplitudes of the forces \citep[computed as in][]{Aubert2017} acting on the fluid core, normalised with the peak of the pressure force and represented as a function of the spherical harmonic degree $\ell$ in states of (a,c) normal and (b,d) reversing polarity of cases (a,b) Stable top 0\% and (c,d) Stable top 29\%. The model times at which force balances have been evaluated are located by vertical dashed lines in Fig. \ref{pathrev}.}
\end{figure*}

The mechanism for dipole attenuation and transition to reversals exposed in Fig. \ref{schematics} only involves the ratio $\tau_\mathrm{exp}/\tau_\mathrm{surf}$ between the time scales of flux expulsion and circulation at the core surface (Fig. \ref{revstats}). It is therefore purely kinematic and independent on the interior dynamics. To substantiate this, we compare two model cases (Fig. \ref{pathrev}a,b) situated at the beginning and at 29\% of a parameter space path leading to the conditions of Earth's core, and featuring a similar value of $\tau_\mathrm{exp}/\tau_\mathrm{surf}\approx 42$. Similar reversals are obtained along this path as $E_\mathrm{mag}/E_\mathrm{kin}$ increases from unity to values exceeding 10 (Fig. \ref{pathrev}c). Unlike previously simulated reversals \citep{Driscoll2009b,Nakagawa2022,TerraNova2024,Frasson2025}, the energy ratio at 29\% of the path also remains high as the dipole vanishes during the event. This means that reversals obtained with the stable top core mechanism occur regardless of the level of inertia relative to the magnetic force. Fig \ref{forcebal} also shows that the leading-order force balance thought to hold in Earth's core between pressure, Coriolis, buoyancy and magnetic forces \citep[the QG-MAC balance,][]{Schwaiger2019} is respected at all times, including during reversals, in addition to being invariant along the path. By controlling and assigning a constant value to $\tau_\mathrm{exp}/\tau_\mathrm{surf}$, we can therefore define a parameter space path towards Earth's core conditions along which the characteristics of polarity reversals are preserved, in addition to the previous invariant features described in \cite{Aubert2017}. The reversal mechanism and the behaviours observed in Fig. \ref{gnmearth}-\ref{skewness} should therefore carry over to the physical conditions of Earth's core.

\section{\label{discu}Discussion}
\subsection{A kinematic mechanism for polarity reversals}
Our results show that when buoyancy sources are located at the bottom of the core, the long-term dipole amplitude is controlled by a competition between subsurface upwelling and surface circulation (Figs. \ref{schematics},\ref{revstats}). To obtain polarity reversals with a large number of Earth-like characteristics (Figs. \ref{gnmearth},\ref{PSVindex},\ref{compliance},\ref{skewness}), we have decreased the mean dipole by decreasing the subsurface upwelling, while using high, but Earth-like values of the magnetic Reynolds number $Rm$ controlling the fluctuations. Doing so, the leading-order interior force balance has remained unchanged (Fig. \ref{forcebal}), which substantiates our claim that the 'stable top core' reversal mechanism exhibited here is purely of kinematic nature. 

This kinematic mechanism fundamentally differs from the dynamical, 'forcing-driven' or 'inertial' mechanism at the base of most previous interpretations of geomagnetic reversals \citep[e.g.][]{Kutzner2002,ChristensenAubert2006,Driscoll2009b,Wicht2010SSR,Christensen2011,Olson2014,TerraNova2024}. This earlier approach exploits a change in force balance and in the system dynamics to cause the transition from stable to reversing dynamos. One of its lesser-known aspects is documented in our study: when approaching the conditions of Earth's core by lowering the magnetic Ekman number $E_\eta$, it becomes gradually more difficult to reach the transition to these 'forcing-driven' reversals. In particular, our results confirm that when $E_\eta<6\times\te{-6}$, the dimensional power $P$ and the magnetic variation time scale $\tau_\mathrm{SV}^{1}$ become unrealistically high and low, respectively, before polarity reversals can be achieved by increasing the forcing. This issue has been previously alluded to by \cite{Christensen2011} from a different angle, by mentioning that the inertial interpretation of the reversal onset required unrealistically small length scales to have a macroscopic influence on the system at Earth's core conditions. \cite{Tassin2021} also noted that in most existing dynamo models, the inertial transition requires kinetic to magnetic energy ratios approaching $E_\mathrm{kin}/E_\mathrm{mag}=1$, a value that is clearly unrealistic at the conditions of the core, as stated in section \ref{intro}. Kinematic reversal mechanisms as the one found here overcome these first-order shortcomings of previous reversing dynamo models (Figs. \ref{pathrev},\ref{forcebal}) and are thus applicable at Earth's core conditions. 

The search for new reversal mechanisms also aims at overcoming the incomplete adherence of the simulated output to paleomagnetic observations. The classical forcing-driven reversals explored by \cite{Sprain2019} could only satisfy three of the five paleomagnetic criteria that they introduced. This number was brought to four in the study of \cite{Meduri2021}, but only within a very narrow window for the forcing parameter. \cite{Jones2025} do not report on adherence to paleomagnetic criteria, but their parameter space survey shows that fine-tuning of the forcing also appears needed to obtain an Earth-like ratio of axial to non-axial dipole. The new mechanism presented here decouples the occurrence of reversals from the level of forcing, thus enabling a reproduction of the full geomagnetic spectrum between secular variations and reversals.  While this is certainly a progress, it should be acknowledged that it currently does not do much better than \cite{Meduri2021} in matching the paleomagnetic criteria, as three of our models only reach $Q_\mathrm{PM}=4$, though several, including that presented in Fig. \ref{gnmearth}, come close to satisfying all criteria. Fine-tuning is also involved here to obtain the correct reversal rate, although it now concerns the level of stratification. We acknowledge that the necessity of fine-tuning does decrease the plausibility of any mechanism, and that our current results have not solved this problem yet.

A successful simulation of geomagnetic reversals should comply with the paleomagnetic field by satisfying all quality criteria. In stable dipole periods, it should furthermore comply with the geomagnetic field by satisfying the morphological criteria and by reproducing a wide spectrum of geomagnetic variations. Finally, all these properties should be obtained while operating in a realistic, plausible and not excessively narrow range of physical conditions. In our study, we have approached this goal by controlling the mean level of the dipole via an additional parameter, while keeping the fluctuations essentially the same. We note that \cite{Jones2025} have arrived to similar results by controlling the level of fluctuations while essentially keeping the mean dipole the same. The two studies exploit the same idea: a purely statistical view of reversals may be adopted, with the underlying dynamical system primarily consisting in the induction equation, and the Navier-Stokes equation only being viewed as a provider of generic random forcing. This view also bridges the gap between self-consistent magnetohydrodynamic models and low-dimensional stochastic approaches to dipole variations \citep{Petrelis2009,Buffett2024}. It is therefore possible that merging our approach with that of \cite{Jones2025} could provide a way to overcome all the above challenges and improve our understanding of geomagnetic reversals.

\subsection{Geodynamic implications}
Our models indicate that a strongly stable top core destroys the magnetic dipole. To obtain a behaviour similar to Earth at present i.e. a dipole-dominated field presenting occasional reversals, the maximum admissible strength of stratification is $N\approx \te{-5}\ut{s^{-1}}$ for layers $H=10\ut{km}$ thick, with even thicker, $H=140\ut{km}$ layers admitting only about a tenth of this value (supplementary Table 2). This is in line with recent simulation results focusing on the reproduction of shorter-term (interannual to decadal) magnetic variations \citep{Aubert2025}. Both these and the present results stand in stark contrast with earlier interpretations of these variations in terms of hydromagnetic waves advocating for strong stratification \citep[$N\approx7\times\te{-5}\ut{s^{-1}}$, $H=140\ut{km}$, e.g.][and following studies]{Buffett2014}. It is also clear that the dynamo mechanism presented here is not compatible with a seismically-inferred density stratification $N\approx\te{-3}\ut{s^{-1}}$ extending 300 km beneath the core \citep{Helffrich2010}, as this would permanently suppress the dipolar component of the field. 

Top core stratification may primarily be caused by a core surface heat flow $Q_\mathrm{CMB}$ below the adiabatic value $Q_\mathrm{ad}$. The adverse density anomaly gradient $\partial C/\partial r$ of equation (\ref{Cgradient}) then has a thermal origin, with a corresponding core surface temperature anomaly gradient $\partial T/\partial r (r_{o})=-N^{2}/\alpha g_{o}$. This leads to the relationship
\begin{equation}
Q_\mathrm{ad}-Q_\mathrm{CMB} = 4\pi r_{o}^{2} k N^{2}/\alpha g_{o}
\end{equation}
Using $k=100\ut{W.m^{-1}.K^{-1}}$ and parameter values as in appendix \ref{thermomodel}, the condition for an Earth-like dipole $N<\te{-5}\ut{s^{-1}}$ translates into a nearly adiabatic core heat flow with $Q_\mathrm{ad}-Q_\mathrm{CMB}<\te{-2}\ut{TW}$. This result implies that virtually any sub-adiabatic heat flow will cause the destruction of the dipole. \rev{Subadiabatic deviations may remain allowed only at the condition that another destabilising source of buoyancy \citep[for instance exsolution buoyancy coming from chemical core-mantle exchanges, e.g.][]{Badro2018} is able to compensate for the stable thermal gradient.} This also means that small fluctuations of $Q_\mathrm{CMB}$ around $Q_\mathrm{ad}$ could have been responsible for the alternation between superchron periods presenting a strong dipole and no reversals, and periods presenting a faint dipole together with an apparent reversal hyperactivity \citep[as documented e.g. in][]{Gallet2016,Gallet2019,Domeier2023}. Rather unexpectedly, in our models an increase of core heat flow causes a decrease in reversal rate. This opposes the trend documented in previous studies, where a larger core heat flow caused more frequent reversals \citep{Kutzner2002,Driscoll2009b,Olson2014}. This shows that several interpretations are in fact possible when investigating the links between changes in reversal rate during the Phanerozoic and the activity of Earth's mantle plumes and subduction flux \citep{Larson1991,Biggin2012,Hounslow2018,Besse2025}. At the very least, reassessments of these links seem warranted once a more complete view of kinematic reversal mechanisms emerges.

Let us examine whether a thermal history can be conceived where, since the nucleation of the inner core, $Q_\mathrm{CMB}$ has remained in the vicinity of $Q_\mathrm{ad}$ and the stable top core reversal mechanism has been operating. Calibrating a thermal evolution model (appendix \ref{thermomodel}) with the convective power $P\approx 2.5-3.4\ut{TW}$ needed in the simulations and in \cite{Aubert2023} to match the observed time scale $\tau_\mathrm{SV}^{1}\approx 400\ut{yr}$ of present geomagnetic variations, equations (\ref{FtoP}-\ref{growthrate}) yield a range $\tau_{i}=500-700\ut{Ma}$ for the age of the inner core. Mantle evolution models constrained by the history of plate tectonics \citep{Olson2013,Choblet2016,Dannberg2024} predict weak variations of the core-mantle boundary heat flow $Q_\mathrm{CMB}$ during this time. The decrease of the adiabatic heat flow $Q_\mathrm{ad}$ has also probably been less than a TW \citep{Labrosse2015}. Assuming that $Q_\mathrm{CMB}$ and $Q_\mathrm{ad}$ have remained constant and, as required by the mechanism, close to each other since inner core nucleation, and neglecting core radiogenic heating \citep{Labrosse2015,Frost2022}, a common present value $Q_\mathrm{CMB}\approx Q_\mathrm{ad}=13-18\ut{TW}$ of these two heat flows is found from equation (\ref{Ei}), in line with estimates derived from the mantle side \citep{Frost2022}. The corresponding core surface thermal conductivity obtained from equation (\ref{Qad}) lies in the range $k=90-120\ut{W.m^{-1}.K^{-1}}$, showing that our initial high conductivity hypothesis is consistent with this thermal history. With our current knowledge, the hypothesis that $Q_\mathrm{CMB}\approx Q_\mathrm{ad}$ since inner core nucleation is therefore plausible, and, as stated above, small fluctuations could explain the alternation of superchrons and reversal hyperactivity periods. This does however not remove the problem that the transition between no reversals and a constantly reversing (and vanishing) dipole occurs within a very narrow core heat flow range, such that the presently observed intermediate reversal rate would require a rather extreme level of fine tuning of $Q_\mathrm{CMB}$ in the vicinity of $Q_\mathrm{ad}$. The scenario, while plausible, is therefore not necessarily the most probable.

Another option is that it is the lateral variations of the core-mantle boundary heat flow, rather than its surficial average, that may have caused the changes in reversal frequency. Adding a pattern derived from a thermal interpretation of lower mantle seismic tomography to a globally adiabatic surface heat flow (models with label Tom in the supplementary Tables), we created convectively stable equatorial regions beneath hot low-shear seismic velocity provinces at the base of the mantle \citep[similarly to][]{Mound2019,Mound2023}. We found upwelling, dipole attenuation levels, and reversal rates consistent with the kinematic mechanism exhibited for homogeneous cases (Fig. \ref{revstats}), showing that the stable top core mechanism operates regardless of the way the top of the core is stabilised. A reversal frequency consistent with Earth's present rate has been found with a peak-to-peak heat flow heterogeneity of order $\Delta q \approx200\ut{mW.m^{-2}}$ which, albeit strong, is in line with earlier determinations from mantle dynamics \citep{Olson2013,Choblet2016,Dannberg2024}. In this case, the reversal rate varies slowly with $\Delta q$, such that the need for fine-tuning is somewhat alleviated. Again in stark contrast with previous studies \citep{Mound2023,TerraNova2024} where reversals were suppressed as the outermost core became partly stable, it is important to stress that here their frequency increases with the level of regional stability (Fig. \ref{revstats}). 

The geodynamo has also featured polarity reversals prior to the expected nucleation time of the inner core \citep{Brenner2022}. A superadiabatic core surface heat flow was in principle required at that time \citep{Labrosse2015,Nimmo2015evolution}, leading to a situation unfavorable to the stable top core mechanism. Still, because reversals in a fully convective sphere remain to be systematically documented by numerical models, future surveys should also offer the possibility to explore and uncover new kinematic mechanisms applicable to Earth prior to inner core nucleation, aside from the classical forcing-driven approach to reversals. 

\section*{Acknowledgements}
\rev{The authors thank two anonymous referees for their comments}. This project has been funded by ESA in the framework of EO Science for Society, through contract 4000127193/\-19/NL/IA (SWARM + 4D Deep Earth: Core). Numerical computations were performed at S-CAPAD, IPGP and using HPC resources from GENCI-TGCC and, GENCI-IDRIS and GENCI-CINES (Grant numbers A014\-0402122 and A0160402122). The authors would like to thank the Isaac Newton Institute for Mathematical Science, Cambridge, for support and hospitality during the programme DYT2 where work on this paper was undertaken. This work was supported by EPSRC grant number EP/R014604/1. The results presented in this work rely on data collected at magnetic observatories. The national institutes that support them and INTERMAGNET are thanked for promoting high standards of magnetic observatory practice (www.intermagnet.org).

\section*{Code and Data availability}
The numerical code is available from the corresponding author upon reasonable request. Data supporting the findings of this study are available together with analysis codes within the supplementary Tables, at \href{https://doi.org/10.18715/IPGP.2024.m0gw7hg5}{doi: 10.18715/ IPGP.2024.m0gw7hg5}, and from the corresponding author upon request.

\appendix
\section{\label{thermomodel}Core evolution model}
Here we outline a simple model relating the present-day convective power $P$ of the dynamo to the inner-core age $\tau_{i}$. Assuming that the core is well-mixed and the core-mantle boundary heat flow is close to adiabatic, the mass anomaly flux $F$ released at the inner-core boundary relates to $P$ through \citep{Buffett1996,Lister2003,Aubert2009}:
\begin{equation}
F\approx P \left(\overline{\psi}-\psi(r_{i})\right)^{-1},\label{FtoP}
\end{equation}
where $\psi=g_{o}r^{2}/2r_{o}$ is the gravitational potential and $\overline{\psi}$ its average over the outer core. The corresponding present-day growth rate of the inner core is \citep{Lister2003,Labrosse2015}:
\begin{equation}
\ddt{r_{i}}{t}= \dfrac{F}{4\pi r_i^2} \left(\Delta\rho_{i}+\alpha\rho_{i} T(r_{i})\dfrac{\Delta S}{C_p}\right)^{-1}.\label{massflux}
\end{equation}
Here $\rho_{i}$, $\Delta\rho_{i}$ and  $T(r_{i})$ are respectively the density, compositional density jump, and temperature at the inner core boundary, $\Delta S$ is the entropy of crystallisation, $C_{p}$ and $\alpha$ are respectively the specific heat and thermal expansion coefficient of the outer core. The age of the inner core is well approximated by \citep{Labrosse2015}:
\begin{equation}
\tau_{i}\approx 0.4 r_{i} \left(\ddt{r_{i}}{t}\right)^{-1}.\label{growthrate}
\end{equation}
We use the following values for physical constants: $\rho_{i}=12200\ut{kg.m^{-3}}$, $r_{i}=1220\ut{km}$, $r_{o}=3485\ut{km}$, from which follows $\overline{\psi}-\psi(r_{i})=8.73\times\te{6}\ut{m^{2}.s^{-2}}$ \citep{Aubert2009},  and values from \cite{Labrosse2015}: $\alpha=\te{-5}\ut{K^{-1}}$, $\Delta\rho_{i}=580\ut{kg.m^{-3}}$, $T(r_{i})=5500\ut{K}$, $\Delta S=127 \ut{J.K^{-1}.kg^{-1}}$, $C_{p}=750\ut{J.K^{-1}.kg^{-1}}$. 

We further relate the inner core age $\tau_{i}$ to the present-day core-mantle boundary heat flow $Q_\mathrm{CMB}$ by assuming that this heat flow has remained constant since the nucleation of the inner core. Neglecting radiogenic heat production in the core leads to the following relationship:
\begin{equation}
Q_\mathrm{CMB}=\dfrac{E_{i}}{\tau_{i}}\label{Ei},
\end{equation}
where $E_{i}$ is the inner core formation energy. We use the value $E_{i}=29.7\times\te{28}\ut{J}$ from \cite{Labrosse2015}. The corresponding thermal conductivity obtained if this core heat flow is exactly adiabatic can be obtained through
\begin{equation}
Q_\mathrm{CMB}=Q_\mathrm{ad}=-4\pi r_{o}^{2}k \,\ddt{T_\mathrm{ad}}{r}\,(r_{o}),\label{Qad}
\end{equation}
where we use the value $\mathrm{d} T_\mathrm{ad}/ \mathrm{d} r\,(r_{o})=-0.97\ut{K.km^{-1}}$ for the present-day adiabatic gradient at the top of the core \citep{Labrosse2015}.

For cases with the Het set-up, equations (\ref{FtoP}-\ref{Ei}) also help to relate the dimensionless peak-to-peak mass anomaly flux heterogeneity $\Delta f/f_{0}$ presented in supplementary Table 1 to a dimensional heat flow heterogeneity $\Delta q$ at the core surface presented in supplementary Table 2. Writing $\Delta q=C_{p} \Delta f/\alpha$, and expanding $f_{0}=F/4\pi r_{o}^{2}$ leads to
\begin{equation}
\dfrac{\Delta q}{q_{0}} = \gamma \dfrac{\Delta f}{f_{0}},
\end{equation}
where $q_{0}=Q_\mathrm{CMB}/4\pi r_{o}^{2}$ and
\begin{equation}
\gamma=\dfrac{0.4\cdot 4 \pi r_{i}^{3}}{E_{i}} \left(\dfrac{C_{p}\Delta\rho_{i}}{\alpha}+ \rho_{i} T(r_{i})\Delta S\right).
\end{equation}
In the main text we use the value $\gamma=1.6$ obtained with the parameter values mentioned above. The core heat flow $Q_\mathrm{CMB}$ is obtained from the dynamo power through the classical efficiency relationship $Q_\mathrm{CMB}=P/\epsilon$, where $\epsilon=\gamma \alpha \left(\overline{\psi}-\psi(r_{i})\right) /C_{p}=0.19$.

\bibliographystyle{cas-model2-names.bst}
\bibliography{Biblio}

\end{document}